
%
\input phyzzx.tex
\tolerance=5000000
\overfullrule=0pt
\pubnum={CERN-TH.7273/94}
\date={May 1994}
\pubtype={}
\titlepage

\rightline{hep-th/9405xxx\phantom{000}}

\title{\bf Effects of Virtual Monopoles}

\author{A. De R\'ujula}
\centerline{  CERN, 1211 Geneva 23 }

\centerline{Switzerland}

\abstract{Electromagnetism would be a ``more unified'' theory if
there were elementary magnetic monopoles and/or particles with both
electric and magnetic charges (dyons). I discuss the simplest
possibilities for the addition of these entities onto the Standard
Model, and their empirical consequences. Lower limits on the masses
of monopoles and dyons stemming from their quantum effects on current
observables turn out to be much stronger than the existing limits
from direct searches. Anomalies in the three-photon decay of the $Z$
constitute good specific signatures for monopoles or dyons. $T$-odd
observables in the $e^+e^-\!\rightarrow\! W^+W^-$ process are
signatures for dyons, but they are severely constrained by existing
data. The subjects of monopolium, monopole cosmology and
non-elementary monopoles are also discussed.}

\vskip 5cm

\noindent{CERN-TH.7273/94}

\noindent{May 1994}
\endpage
\sequentialequations
\pageno=1

\REF\rDirac{P.A.M. Dirac, Proc. Roy. Soc.

(London) Ser. $\bf A133$ (1931) 60.}
\REF\rSchwinger{J. Schwinger,
      Phys. Rev. {\bf 151} (1966) 1048, 1055.}
\REF\rtP{G. 't Hooft, Nucl. Phys. $\bf B79$ (1974) 276; \hfill \break
A.M. Polyakov, JETP Lett. $\bf 20$ (1974) 194.}
\REF\rmoma{E.B. Bogomolny, Soviet J. Nucl. Phys. {\bf 24} (1976)
801;\hfill \break
L.D. Fadeev, Lett. Math. Phys. {\bf 1} (1976) 289;\hfill \break
           S. Coleman {\it et al.}, Phys. Rev. {\bf D15} (1977) 544.}
\REF\rFR{J. Schwinger, Phys. Rev. {\bf 144} (1966) 1087.}

\REF\rWZ{B. Zumino, in {\it Int. School of Physics, Ettore Majorana},
edited by A. Zichichi (Academic, New York, 1966);
\hfill \break S. Weinberg, Phys. Rev. {\bf 138} (1965) B988; \hfill
\break J.C. Taylor, in {\it Lectures in High Energy Physics,} edited
by H.H. Aly (Wiley, New York, 1968); \hfill \break A. Rabl, Phys.
Rev. {\bf 179} (1969) 1363.}

\REF\rZ{D. Zwanziger, Phys. Rev. {\bf D3} (1971) 880, {\bf D6} (1972)
458.}
\REF\rDeans{W. Deans, Nucl. Phys. {\bf B197} (1982) 307.}
\REF\rconf{E. Dagotto and J. Kogut, Nucl. Phys. {\bf B295} (1988)
123; \hfill \break J. Kogut, E. Dagotto and A. Ko\v ci\v c, Phys.
Rev. Lett. {\bf 60} (1988) 722; \hfill \break S. Hands and R.
Wensley, Phys. Rev. Lett. {\bf 63} (1989) 2169.}
\REF\rSTU{D.C. Kennedy and B.W. Lynn, Nucl. Phys. {\bf B322} (1989)
1;\hfill \break
D.C. Kennedy {\it et al.} Nucl. Phys. {\bf B231} (1989) 83;\hfill
\break
B.W. Lynn, Stanford Univ. Preprint SU-ITP-867 (1989);\hfill \break
          M.E. Peskin and T. Takeuchi, Phys. Rev. Lett. {\bf 65}
(1990) 964.}
\REF\rHG{H. Georgi, in {\it Weak Interactions and Modern Particle
Theory} (Benjamin/Cummings Pubs. Menlo Park, 1984).}
\REF\rYW{T.T. Wu and C.N. Yang, Nucl. Phys. {\bf B107} (1976) 365.}
\REF\rNOV{V.A. Novikov, L.B. Okun and M.I. Vysotsky, Nucl. Phys. {\bf
B387}
          (1993) 35.}
\REF\rCDF{F. Abe {\it et al.},
The CDF collaboration, Fermilab-PUB-94/097-E. May 1994.}
\REF\rGSG{S. Graf, A. Sch\"afer and W. Greiner, Phys. Lett. {\bf
B262}
          (1991) 463.}
\REF\rEH{W. Heisenberg and H. Euler, Z. Phys. {\bf 98} (1936) 714,
\hfill \break J. Schwinger, Phys. Rev. {\bf 82} (1951) 664, {\bf 93}
(1954) 615, {\bf 94} (1954) 1362.}
\REF\rstan{E.W.N. Glover and A.G. Morgan, Z. Phys. {\bf 60} (1993)
175.}
\REF\ree{F. Berends and R. Kleiss, Nucl. Phys. {\bf B186} (1981) 22.}
\REF\rGKR{C.J. Goebel, in {\it Quanta, Essays in Theoretical Physics,
Dedicated to Georg Wentzel,} edited by P.G.O. Freund, C.J. Goebel and
Y. Nambu (Chicago, 1970); \hfill \break E. Kyriakopoulos and R.
Ramachandran, Nuovo Cim. Lett. {\bf 15} (1976) 161.}
\REF\rHDR{A. De R\'ujula and H. Georgi, Phys. Rev. {\bf D13} (1975)
1296; \hfill \break E.C. Poggio, H.R. Quinn and S. Weinberg, Phys.
Rev. {\bf D13} (1975) 1958.}
\REF\rLBW{A.C. Longhitano, Phys. Rev. {\bf D22} (1980) 1166 and
Nucl. Phys. {\bf B188} (1981) 118;\hfill \break

T. Appelquist,

in {\it Gauge Theories and Experiments at High Energies}, edited by
K.C. Brower and
D.G. Sutherland, (Scottish University Summer School in Physics Publ.,
St. Andrews
1980); \hfill \break W. Buchm\"uller and D. Wyler, Nucl. Phys. {\bf
B268} (1986)
621.}
\REF\rGavela{A. De R\'ujula, M.B. Gavela, P. Hernandez and E.
Mass\'o, Nucl. Phys. {\bf B384} (1992) 3.}
\REF\rBMDR{A. Bilal, E.
Mass\'o and A. De R\'ujula, Nucl. Phys. {\bf B355} (1991) 549.}
\REF\rGuth{A.H. Guth, Phys. Rev. {\bf D23} (1981) 347; \hfill \break
A.D. Linde, Phys. Lett. {\bf 108B} (1982) 389; \hfill \break A.
Albrecht and P.J. Steinhardt, Phys. Rev. Lett. {\bf 48} (1982) 1220.}
\REF\rKT{For a review, see E.W. Kolb and M.S. Turner, {\it The Early
Universe} (Addison-Wesley Pub. Co. New York 1990).}
\FIG\fprop{(a) A one-loop virtual-monopole correction to the
           propagator of neutral vector bosons $V=Z,\,\gamma$.

           (b) A two-loop correction.}

\FIG\flims{Lower limits on $m/n$ as a function of $m_t$,

           with $m$ the monopole mass and $n$ its charge in
elementary
           units, (a) at the 1$\sigma$ confidence level, (b) at
3$\sigma$.
The shaded areas are allowed.           The recently published CDF
result on $m_t$ narrows the allowed domain down to the barred
regions.}

\FIG\floop{$Z$-production in $e^+e^-$ annihilation and
monopole-induced
           $Z\rightarrow 3\gamma$ decay;

           (a) with the dual photon vertices (denoted by a heavy dot)
               hooked to the monopole line and the standard
               vertex on the electron line.

           (b) A dual viceversa.
           (c) A higher-order correction to (a).
           (d) The QED background.}
\FIG\fwidth{Branching ratio for monopole-induced $Z\rightarrow
3\gamma$
            decay, as a function of $m/n$, in the notation of
Fig.$\,$\flims.}
\FIG\fleptwo{Corrections to the $VW^+W^-$ vertex, with
$V=Z,\,\gamma$. (a) A lepton loop. (b,c) Two dyon loops. Only the
heavy dots are magnetic couplings.}

\noindent{\caps 1. Introduction}

The quest for symmetries in the laws of Nature, sometimes exact,
occasionally broken in more or less elegant ways, has been crowned by
a long
list of successes. This, and not only aesthetical pleasure, is a good
reason to pursue the quest.
Often quoted as the first, or second, ``grand unification'' is that
of
electricity and magnetism. But Maxwell's or Nature's feat is
only partial: the conceivable symmetry between electric and magnetic
charges
is broken by the non-existence of (relatively light) magnetic
monopoles.

A higher symmetry was one of Dirac's motivations
to introduce the notion of
magnetic charges [\rDirac].
But he found that the elementary electric and magnetic
charges $e$ and $g$
ought to be quantified so that $e\,g=2\,\pi\,n$, with $n$
an integer (units are such that
$\alpha_e\equiv e^2/4\,\pi\sim 1/137$,
$\alpha_g\equiv g^2/4\,\pi\sim 34\, n^2$). Thus, in a sense, the
addition of
Dirac's monopoles to QED would not constitute a truly ``grand''
unification of
electricity and magnetism, since $e\neq g$. Perhaps this asymmetry
can be cured as
in conventional Grand-Unified Theories, by recalling that, in a
quantum
theory, couplings are energy-dependent and it is only at a scale
wherein masses are negligible that the couplings merge into a single
value (the Dirac condition $e(q^2)\,g(q^2)=2\,\pi\,n$ has been argued
to hold [\rSchwinger] for the
one-loop renormalized quantities at any given momentum scale $q^2$,
so that as $e$ increases with $q^2$, $e$ and $g$ do tend to merge).

Interest in monopoles revived in the 70's, following the discovery by
't
Hooft and Polyakov [\rtP] that there exist monopole solutions
to the field equations of theories in which a semi-simple unifying
gauge group
is broken to the $U(1)$ of QED. If the unification scale is $M_{_{\rm
GUT}}$,
the monopole mass [\rmoma]
is $m\ge M_{_{\rm GUT}}/\bar{\alpha}$, with $\bar{\alpha}$ the
grand-unified
fine-structure constant. A consequence of these theoretical
developments
is that the feeling has permeated the community that monopoles are
extremely heavy
objects, desperately out of the reach of accelerator experiments. I
perceive this circumstance as negative and I contend that monopoles
much lighter than the
``grand-unified'' ones may simply {\it happen} to exist, and should
be looked for.

Almost
 everywhere, I assume monopoles
to be spin-$1/2$ ``point-like''
particles whose radius, $R$, is negligible
relative to the inverse of their
mass, in contrast to grand-unified monopoles,
extensive
objects with $R\sim 1/(m\bar{\alpha})$. In Section 7, I discuss
monopoles that are not point-like, but are unconventional in being
much lighter than the grand-unified ones.

The extension of QED to a theory containing both electric and
magnetic
elementary
charges is non-trivial and, in the case of point-like monopoles,
it has not reached the theoretically satisfactory point wherein
consistency and
renormalizability have been demonstrated to all orders
of perturbation theory (even if one were to finesse the problem that
a
perturbative expansion cannot be simultaneously convergent in $e$ and
$g$).
The Feynman rules have been derived [\rFR--\rDeans] and are
consistently usable at the one-loop level, in the case of monopoles.
For dyons the situation is more obscure (these points are discussed
in Section 2).

For the purposes of this paper, to estimate the current limits on
monopole
masses and to suggest concrete strategies for indirect monopole
searches,
our current understanding of monopole theory is (almost) all one
needs.

The standard $SU(2)\otimes U(1)_{_Y}$ model is sufficiently well
established for
it to be necessary, when considering monopoles, to discuss the
gauge-group representation to which they might belong. To avoid
anomalies
and other potential strictures, one is most economically led to
introduce ``vector-like''
spin-$1/2$ monopoles, whose left- and right-handed projections belong
to
the same group representation. I consider only the two simplest
cases, that of electrically neutral $SU(2)$-singlet monopoles
$M^0_{_{L,R}}$
and that of ``dyons'' $(D^0,D^-)_{_{L,R}}$, doublets with
$(\nu^0,e^-)$-like electrical charges\foot{I refrain here from
discussing representations containing objects of different magnetic
charge; this would lead to the introduction of magnetically-charged
gauge bosons
and an enlarged gauge group.}. In what follows I
refer to these two choices simply as monopoles and dyons,
or to both as monopoles when no distinction is necessary.

In both cases the magnetic charge is commensurate \`a la Dirac:
$g=2\,\pi\,n/e$. To economize parameters, I generally let the dyons
be
(quasi)-degenerate, $|m(D^-)-m(D^0)|\ll m(D^i)$.

The current to which the standard
$Z^0$ boson couples has an electromagnetic component, so that not
only
photons, but also $Z$'s must couple to monopoles, with a large
amplitude of
$O(g)$. If monopoles are not confined, as they might [\rconf]
by the strong monopole--antimonopole
forces, the fact that LEP detectors have
 not been swamped by monopoles immediately
implies a ``direct'' limit $m>M_{_Z}/2$.

We show in Section 3 how the trumpeted success of the
Standard Model in accommodating all current data implies
a lower limit on the monopole mass of order 1 TeV, a much stronger
result than the direct one\foot{This is so in spite of the fact that,
with our choice of monopole properties, the ``oblique'' vector-boson
propagator
corrections $S$, $T$ and $U$, at least in their original disguise
[\rSTU], vanish
automatically.}. In Section 4 we discuss how the  $Z$ decays into odd
numbers of photons constitute a similarly sensitive but much more
specific
limiting (or discovery) test.
The masses of our dyons are constrained by current data
in the same way as those of monopoles; in Section 6 we discuss how
dyons,
which couple not only to $\gamma$'s and $Z$'s, but also to $W$'s, may
induce very characteristic $T$-odd observables in the
$e^+e^-\rightarrow W^+W^-$ process. But the available data constrain
these effects below the level of observability at LEP-II. I do not
discuss monopole signatures at
higher-energy $e^+e^-$ colliders.

The monopole charge is too  large for
a perturbative expansion in $g$ to be trustworthy. This unsurmounted
caveat will unavoidably haunt all our numerical considerations: an
argument
is needed to judge the credibility of the results.

Throughout the paper, we discuss how monopoles may modify
the standard relations between observables measured at energy scales
below the monopole mass, $q^2<m^2$. The incidence of monopoles can
thus be characterized in
the customary fashion, by grafting onto the standard Lagrangian a
series of
dimension $d>4$ effective operators induced by virtual
monopoles, whose degrees of freedom have been integrated out, or,
more
pictorially, integrated ``in''. In each case we characterize the
effect of monopoles by the lowest-dimension relevant operator (this
is
tantamount to keeping the first non-vanishing term in an expansion in
$q^2/m^2$).
All we assume is that the {\it order of magnitude} of the coefficient
of each of these operators can be estimated by computing it to lowest
non-vanishing order in
$g$. This is indeed an assumption: in every case the corrections
(to the next order in $g$) are $O(g^2/16\,\pi^2)\sim 1$.

Equivalent to the above is the approach used
in other strongly-interacting realms, such as the chiral Lagrangians
describing the low-energy interactions of pseudoscalar mesons. There,
the size of the coefficient of a particular operator is set by
demanding
that its estimates to successive orders of perturbation give results
of the same magnitude [\rHG]. The coefficients of the effective
operators
we deal with involve powers of $g/m$. In a ``chiral-like'' approach
one would
simply substitute a multiple of $g/m$ by the analogue of
$f_\pi^{-1}$.
In a sense, computing diagrams explicitly, as we do, is simply a way
to keep straight
the powers of $4\,\pi$.

Monopole--antimonopole bound states may occur below the threshold for
``open'' monopole production, and need to be discussed. I argue in
Section 5 that these states are so wide that they should not be
considered explicitly. Their collective effects should already be
included
in the ``partonic'' description in terms of virtual-monopole loops.

Arbitrarily heavy elementary monopoles or dyons would antagonize the
standard cosmological lore. The ensuing upper mass limits are
discussed in Section 8. A summary and the conclusions are offered
in Section 9.

Since the subject of monopoles has not been investigated at length in
the spirit of this paper, I have taken the liberty to report on the
results of the exploration of various alleys which do not take to
fertile destinations. Consequently, a positive-thinking reader may
consider skipping Sections 5 to 7, which deal with monopolia,
$W$-pair production and extended monopoles, and to concentrate on
Section 4, that deals with the phenomenologically relevant subject of
the multiphoton decays of the $Z$.

\vskip .2cm
\noindent{\caps 2. Electrodynamics with electric and magnetic poles}

It is not possible to write a theory of point-like electric {\it and}
magnetic charges wherein the electromagnetic field is described
exclusively by a local vector potential $A^\mu(x)$. Whence the
necessity of introducing the Dirac string [\rDirac]
or a multivalued potential [\rYW]. The string, or its surrogates, can
be regarded as gauge-artefacts and are unobservable if the Dirac
charge-quantization condition is satisfied. Yet, they entail a
measure of non-locality in the action and in probability amplitudes,
that must not make its way to the observables.

An action describing electric and  magnetic poles can be translated
with the customary methods into a set of Feynman rules. Very little
attention has been paid to these rules, the conventional wisdom (that
I have challenged in the Introduction) being that the large value of
$g$ makes them entirely useless. These calculational rules are of
delicate handling in the case of monopoles, and are worse than
delicate for dyons.

Let $n$ be a space-like four-vector tangent to the Dirac string,
$\epsilon_{\mu\nu\rho\sigma}$ the fully antisymmetric four-index
symbol, and $\epsilon^\rho_{_A}$ a photon-polarization vector. The
monopole counterparts to the ${\cal L}_e\! =\! -i\,e\, A\cdot J$
coupling of a photon to a charged object, the configuration- and
momentum-space vertex ``Feynman rules'' for the coupling to a
monopole, are [\rWZ--\rDeans]:
$$
{\cal L}_g=-\,i\,g\,{\epsilon \; (n,\partial_x ,A(x),J(x))\over n
\cdot \partial_x}
\leftrightarrow
 -\,i\,g\,
{\epsilon_{\mu\nu\rho\sigma}\,
n^\mu q^\nu \epsilon^\rho_{_A}\,
J^\sigma
\over n \cdot q + i \, \epsilon}\; ,
\eqn\etrouble
$$
with $J_\sigma\! = \!
(p_\sigma+p'_\sigma)$ or $\bar{\psi}(p')\gamma_\sigma\psi(p)$, for
spinless and spin-$1/2$ monopoles.
The $\epsilon$ tensor is needed to describe a ``dual'' coupling, and
the vector $n$ must be contracted with it, since there are only two
independent four-vectors in the set $q,p,p'$. The terrifying
denominators in Eqs.$\,$\etrouble\ embody  the non-local character of
the Dirac string.

Our $M^0_{_{L,R}}$ monopoles and $(D^0,D^-)_{_{L,R}}$ dyons are
``magnetic'' weak-$SU(2)$ singlets, but they carry $U(1)_{_Y}$
magnetic charge. Their couplings to $Z$'s and $\gamma$'s are the dual
or magnetic counterparts to the standard couplings of a singlet with
non-vanishing hypercharge:
$$
{\cal L}_{_D}=-i\,\left({g\over c}\right)
\,{\epsilon \; (n,\partial_x ,B(x),J(x))\over n \cdot \partial_x}\; ,
\eqn\edyco
$$
where
$$
B_\mu\equiv c\,A_\mu - s\,Z_\mu \; ,
\eqn\eGinv
$$
is the $U(1)_{_Y}$  potential, and
$s,c=\sin(\theta_{_W}),\cos(\theta_{_W})$.
Our electrically-neutral monopoles, $M^0$, have no gauge couplings
other than the ones of Eq.$\,$\edyco, while dyons also sport standard
``electric'' vector-like couplings to $W$'s, to $Z$'s and, in the
case of $D^-$, to $\gamma$'s.

To illustrate the use of Eq.$\,$\etrouble, consider the
photon-mediated processes of $e^+e^-$ annihilation into lepton ($L$),
monopole ($M$) or dyon ($D$) pairs.
The matrix element squared for $L^+L^-$ production is of the form
$|J_e^\dagger\cdot J_{_L}|^2/q^4$. The calculation of the cross
section for $e^+e^-\rightarrow M \bar{M}$ production with casual use
of the rule \etrouble\ leads to an $n$-dependent result, even after
the erasure of all terms that vanish because of current conservation
($q\cdot J_e\! =\! q\cdot J_{_M}\! =\! 0$). The somewhat debonair
recipe [\rDeans] of dropping all terms proportional to $q^2$ (that
would cancel the photon's propagator pole) results in a squared
matrix element proportional to
$(|J_e^\dagger\cdot J_{_M}|^2-
|J_e|^2\, |J_{_M}|^2)/q^4$, a sensible answer\foot{The crossed
process of electron scattering on a monopole has the same squared
matrix element, from which one can reproduce the classical results
for the scattering on a static monopole of electrons that do not hit
the string.} at last. The total cross sections for the production of
magnetically and electrically charged pairs of a given spin have the
same functional form, as expected; while their angular distributions
are different, also as expected.

So far, so good. The troubles arise as one attempts to extend these
considerations to the $e^+e^-$ production of dyon pairs. The Feynman
rules are simply obtained [\rZ] by adding to Eq.$\,$\etrouble\ a
conventional electric-charge coupling of the $D^-$ dyon. The
interference term between the ``electric'' and ``magnetic'' dyon
amplitudes does not cancel in the expression for the cross section
and is proportional to $\epsilon\, (n,p,p',k)$ with $k$ one of the
incoming momenta and $n$, alas, the vector defining the Dirac string.
I have neither solved this problem nor found it discussed in the
literature. Clearly, the bugaboo is in the treatment of superimposed
singularities at ``the end of the string''.  Presumably a study of
the effective vertex operators describing the long-wavelength limit
of the photon couplings to (not point-like) topological dyons would
help clarify these issues.

The slippery character of all these grounds emanates from the
necessity of using vector potentials, rather than $\vec E$ and $\vec
B$ fields, in describing photonic couplings to monopolar electric and
magnetic charges. There is no difficulty in translating a
magnetic-dipole coupling such as $F_{\mu\nu}
\bar{\psi}\sigma^{\mu\nu}\psi$ into its dual electric-dipole
counterpart, suffice it to trade $F_{\mu\nu}$ for
$\tilde{F}_{\mu\nu}\equiv {1\over 2}
\epsilon_{\mu\nu\rho\sigma}F^{\mu\nu}$. This simple remark will allow
us, in discusssing dyons in Section 6, to bypass  the problem of
dealing with inconsistent Feynman rules.

\vskip .5cm
\noindent{\caps 3. Current limits on the masses of monopoles and
dyons}

It has become a flourishing industry to set limits on novel
effects by exploiting the
accurate success of the Standard Model in relating precise
electroweak
data at energies up to $M_{_Z}$ (see [\rNOV] for a clear review).
Typically,
three of the best measured quantities ($\alpha$, $M_{_Z}$ and the
Fermi
constant as extracted from muon beta decay) are used as inputs to
specify the parameters and predict other quantities.
The predictions depend significantly on the mass of the top quark,
$m_t$, and to a lesser extent on the Higgs boson mass, $M_{_H}$.

The CDF group has announced [\rCDF] evidence for the $t$ quark, at
$m_t=175\pm 15$ GeV, a result that we take into account where
appropriate. The uncertainty in
$M_{_H}$ can be conveniently dealt with by allowing it to vary from
its
experimental lower limit ($\sim 50$ GeV) to some theoretically
sensible
upper limit ($\sim 1$ TeV) and adding the effect linearly as a
``theoretical
error'' in the prediction at hand. Comparison with measured values of
the predicted quantities then results in limits on the parameters of
some
non-standard effect, as functions of $m_t$. In dealing with limits on
the
mass of monopoles, we follow this traditional procedure.

The most relevant predicted observables in constraining monopole
masses
turn out to be $g_{_A}$ and
$g_{_V}/g_{_A}$, the vector and axial couplings of the $Z$ to charged
leptons (extracted from the $Z$ leptonic widths and asymmetries) and
$M_{_W}/M_{_Z}$. For
ease of reference, I collect below the standard results (the data and
analysis are as in [\rNOV], and the $m_t$-dependences are my own
simplistic
fits to the full predictions of the Standard Model, sufficiently
accurate
over the relevant $m_t$ range; the $1\sigma$ errors in the
predictions
are given in
parenthesis, the theoretical Higgs-related error is labelled
$[M_{_H}]$):
$$
{g_{_V} \over g_{_A}}=
0.0753(12)+0.00345\,
\left[{m_t^2-(176\,{\rm GeV})^2 \over M_{_Z}^2} \;\pm 1.0\,[M_{_H}]
\right] \; ,
\eqn\egvgast
$$
$$
-g_{_A} = 0.5(0)+0.00065\,
\left[{m_t^2-(111\,{\rm GeV})^2 \over M_{_Z}^2} \;\pm 0.4\,[M_{_H}]
\right] \; ,
\eqn\egammanust
$$
$$
{M_{_W} \over M_{_Z}}=
0.8768(2)+0.00163\,
\left[4.25\,{m_t-100\,{\rm GeV} \over M_{_Z}} \;\pm 0.6\,[M_{_H}]
\right]\; .
\eqn\eMWMZst
$$

These are to be compared with the experimental results:
$$
{g_{_V} \over g_{_A}}=0.0728(28)\;\;\;\;\;
-g_{_A}=0.4999(9)\;\;\;\;\;
{M_{_W} \over M_{_Z}}=0.8798(29) \; .
\eqn\eexp
$$

Let $V=\{\gamma,Z\}$ represent the pair of neutral gauge bosons.
Virtual monopoles contribute to the $V$-propagator via diagrams such
as
those in Fig.$\,$\fprop, and modify the standard predictions for the
observables
mentioned in the previous paragraph. As discussed in the
Introduction,
we estimate the effect of monopoles by focusing on
the pertinent lowest-dimension
effective operator (induced by integrating ``in'' the monopole
degrees
of freedom) and computing its coefficient to leading order in $g^2$.
In the case of the $V$-propagator, this is equivalent to computing,
to
$O(m^{-2})$, the $O(g^2)$ diagram of Fig.$\,$\fprop a and neglecting
corrections of $O(g^2/16\,\pi^2)$ or higher, such as those
represented by the
diagram of Fig.$\,$\fprop b.

The results for the putative
departures (denoted $\Delta$) from the standard predictions, are:
$$
\Delta \,{g_{_V} \over g_{_A}} =
{16\,s^4\,c^2 \over c^2-s^2}\;
{g^2\over 4 \pi^2}\;
{M_{_Z}^2 \over 15 \, m^2}\; ,
\eqn\egvga
$$
$$
\Delta(g_{_A})=
{s^2\over c^2}\;
{g^2\over 4 \pi^2}\;
{M_{_Z}^2 \over 15 \, m^2}\; ,
\eqn\egammanu
$$
$$
\Delta\,{M_{_W} \over M_{_Z}}=
{c\,s^2\over c^2-s^2}\;
 {g^2\over 4 \pi^2}\;
{M_{_Z}^2 \over 15 \, m^2}\; ,
\eqn\eMWMZ
$$
where $s^2\equiv1-c^2\equiv \sin^2 (\theta_{_W}) \sim
0.231$\foot{Notice that
with our conventions for $e$ and $g$ and the corresponding currents,
these corrections are a factor of two
larger than the ones that would be obtained by first
computing the effect of a singlet vector-like heavy lepton $E^-$ of
mass
$m$ and then substituting $e$ for $g$ in the result.}.

The various 1$\sigma$ and 3$\sigma$ limits on the monopole mass $m$
as  functions of
$m_t$ are respectively displayed in Figs.$\,$\flims a  and  \flims b,
for a monopole of minimum magnetic charge,
$n=1$ (since the monopole-induced corrections
``$\Delta$'' are all quadratic in $g/m$, the figure can
also be interpreted as the limits on $m/n$, for a monopole of charge
$g=2\,\pi\,n/e$, with $n\ge 1$).  The combined lower limits are:
$$
m>n\times 1.0\;{\rm TeV, \; at }\; m_t=147\;{\rm GeV}\;\;(1\sigma)\;
,
\eqn\emmina
$$

$$
m>n\times 0.7\;{\rm TeV, \; at }\; m_t=137\;{\rm GeV}\;\;(3\sigma)\;
,
\eqn\emminb
$$

where in the numerics I have used the value of $g$ that
corresponds to $\alpha_e(M_{_Z}^2)\sim 1/128$. As can be seen in
Fig.$\,$\flims, the recent CDF result [\rCDF] on $m_t$ modifies
Eq.$\,$\emmina\ to
$$
m>n\times 1.2\;{\rm TeV, \; at }\; m_t=165\;{\rm GeV}\;\;(1\sigma),
\eqn\emminc
$$
while the $3\sigma$ result of Eq.$\,$\emminb\ stays put.

For our mass-degenerate dyons, the $D^0$ and the $D^-$ loop give
contributions to the various $\Delta$'s identical to those of a
singlet
monopole and the above lower-mass limits are to be multiplied by
$\sqrt{2}$. At first sight, the dealer in the art of setting limits
on physics beyond the Standard Model may be surprised that, in spite
of their strong couplings, monopoles and dyons are not more
constrained than in Eqs.$\,$\emmina--\emminc. The reason for the
relative weakness of these strictures is that vector-like monopoles
and quasi-degenerate dyons are very particularly guileful at avoiding
all low-energy constraints.

Given our perilous use of a
perturbative expansion in $g$, Eqs.$\,$\emmina--\emminc\  are only
estimates of monopole-mass limits.
But these limits are so much larger than $M_{_Z}$ that
it is fair to conclude that {\it direct}
searches for monopoles at LEP are unlikely to be successful\foot{To a
lesser
extent, this grim
conclusion could also have been reached from a previous analysis,
along similar lines, of the monopole contributions to the anomalous
magnetic moment of
the muon [\rGSG], that result in a limit $m>120$ GeV.}.

\vskip .2cm
\noindent{\caps 4. Z decays into multiple photons}

There are ways to search for virtual-monopole effects more specific
than
the ones discussed in the previous Section. Given the very large
couplings
of monopoles to $Z$'s and $\gamma$'s, the most obvious candidate is
the
monopole-mediated decay of
a $Z$ into an odd number of photons, $l$, as depicted in
Fig.$\,$\floop a
for $l=3$.
Because of the naturally large couplings of monopoles to the gauge
bosons, it is difficult to imagine a sensible theory that, for
a fixed scale of its new dynamics, could modify $Z\rightarrow
l\,\gamma$
decays as much as monopoles would.

Even for the dominant ($l=3$) decay, the standard $Z\rightarrow l \,
\gamma$
transition, induced by all real and virtual charged-particle
intermediate states,
has a totally negligible branching ratio $\sim 2.8\times 10^{-10}$
[\rstan].
The main conventional  source of $3\,\gamma$ final states
at $\sqrt{s}=M_{_Z}$ is the pure QED $e^+e^-$ annihilation depicted
in
Fig.$\,$\floop d. This process has a characteristic
photon-bremsstrahlung
behaviour, with its cross section peaking when one of the photons is
soft, or collinear to one of the colliding particles.
For photons sufficiently hard an acollinear to be observable in LEP
detectors,
$\sigma (e^+e^-\rightarrow 3 \, \gamma) / \sigma_{_{TOT}} \sim {\rm
a\, few}\,10^{-6}$ at
$\sqrt{s}=M_{_Z}$
[\ree]. This means that for millions of produced $Z$'s, the QED
process
is marginally observable\foot{There are clearly other
detector-related backgrounds, such as $\pi^0$'s or even jets
masquerading
as single photons, but these are not for a theorist to deal with.}.

Since the standard $e^+e^-\rightarrow 3\gamma$ process is not
negligible,
in studying a possible deviation from the expectations one ought to
compute the interference of the QED amplitude of Fig.$\,$\floop d
with that for a monopole-induced
transition, Fig.$\,$\floop a. Prior to the discovery of a
significant excess of $3\,\gamma$ final states (or of a departure
from the QED-predicted angular and energy distributions), it appears
reasonable
to refrain from this laborious task.
For our current purposes, it suffices to deal with the pure
monopole-induced
$Z\rightarrow l\,\gamma$ partial widths.

We estimate the virtual-monopole mediated
$Z\rightarrow l\,\gamma$ amplitudes by computing to leading order in
$g$ the corresponding lowest-dimension effective operators. This
involves
a calculation of Feynman diagrams such as that of Fig.$\,$\floop a,
and the neglect of higher-order effects like the one illustrated by
Fig.$\,$\floop c.  In the figure we
emphasize the distinction between the conventional $Z$--$e$ vertex
and the dual $\{Z,\gamma\}$-monopole couplings of
Eqs.$\,$\edyco,\eGinv\ by denoting the latter with a dot. The
simplest way to present
the results is to trade the calculation of Fig.$\,$\floop a for that
of
Fig.$\,$\floop b, that is, to use the theory's duality to treat the
electron as a monopole
and the monopole as a conventional QED charged
lepton, while keeping straight the powers of $e$ and $g$ and
substituting
$\bar{\psi}_e\,\gamma_\sigma \,(g_{_V}\! +\! g_{_A}
\,\gamma_5)\,\psi_e$ for $J_\sigma$ in
Eq.$\,$\etrouble.

But for a rescaling of coupling constants, we have reduced our task
to the calculation of the
effective interactions describing the couplings of an even number
($l$+1)
of vector bosons, induced at the one-loop level by a massive charged
spinor
field to which they couple as in QED. This is precisely the
definition
of the Euler--Heisenberg Lagrangian [\rEH] describing
light-by-light scattering (and its generalizations) at scales below
$m_e$.
The effective interactions relevant to monopole-induced
$Z\rightarrow 3\gamma$ and $Z\rightarrow 5\gamma$ transitions are:
$$
{\cal{L}}_3=
{\alpha_g^2 \over 360\,m^4}\; {s \over c} \,
\left\{ 4\,(F \! \cdot \! F)^2+7\,(F \! \cdot \! {\widetilde F} )^2
\right\} \, ,
\eqn\eLthree
$$
$$
{\cal{L}}_5=
{\pi\,\alpha_g^3 \over 22680\,m^8}\; {s \over c}\,
\left\{ 8\,(F \! \cdot \! F)^3+13\,(F \! \cdot \! F)(F \! \cdot \!
{\widetilde F})^2\right\}\; ,
\eqn\eLfive
$$
where $F\!\cdot\! F\! =\! F^{\mu\nu}F_{\mu\nu}$, to be properly
symmetrized over all photons. Notice the factor $s/c=\tan
\theta_{_W}$ arising from the substitution of one photon by a
vectorially-coupled $Z$ in the conventional multiphoton effective
Lagrangian.
Given these results, it is only a matter of some toil to compute
the rates, energy- and angular-distributions of the final-state
photons.

It is convenient to discuss first the total
cross sections $\sigma(e^+e^- \! \rightarrow \! Z \! \rightarrow \!
l\,\gamma)$,
and to re-express them as
$Z$ partial widths. For $l=3$:
$$
\Gamma(Z\rightarrow 3\,\gamma)=M_{_Z} \;
{17\,\pi\,s^2 \over 972000\,c^2}\;
\left[ {g^2 \over 4\, \pi^2} \right]^4\;
\left({M_{_Z} \over m} \right)^8\, ,
\eqn\ethreew
$$
while, for $l=5$\foot{Here an overall factor $1/(2^{30}\,3^3\,\pi^7)$
is phase space, the result for an ${\cal {L}}_5=\phi^6/M_{_Z}^{2}$
coupling between distinguishable scalars. The quantity in curly
brackets is an estimate (of better than 1\% precision) of the average
over the eight-dimensional phase space of the normalized matrix
element squared,  i.e. the result of substituting the six $F$'s
within the curly brackets of Eq.$\,$\eLfive\  by all permutations of
$F_1$...$F_6$, adding, dividing by 6!$M_{_Z}^6$, squaring, and
summing over polarizations; the smallness of this number reflects the
derivative nature of the coupling to the five photons that share the
available energy.}:
$$
\Gamma(Z\rightarrow 5\,\gamma)=M_{_Z} \;
{\pi\,s^2 \over 2^{25}\,3^{11}\,5^2\,7^2\,c^2}\;
\left[ {g^2 \over 4\, \pi^2} \right]^6\;
\left({M_{_Z} \over m} \right)^{16}\;\{\, 3.92\times 10^{-4} \, \}\,
{}.
\eqn\efivew
$$

Notice that the r.h.s$.$ of Eq.$\,$\ethreew\ is a function of $m/n$,
with
$n$ the monopole charge in elementary units, and recall that the
other monopole-induced effects of Eqs.$\,$\egvga--\eMWMZ\
also depend on $m/n$. Thus, in a search for virtual-monopole
signatures, the relative
merits of the indirect limits of Section 3 and of the more specific
$Z\rightarrow 3\gamma$ rate are independent of $n$.

Figure \fwidth\ shows Eq.$\,$\ethreew\ converted into a branching
ratio with use of
$\Gamma_{_Z}^{^{\rm TOT}}=2.487$ GeV, and plotted as a function of
$m/n$.
An anomaly in this branching ratio of $O(10^{-6})$ is currently
observable at LEP-I, and
corresponds to $m=n\times 0.60$ TeV. This mass value is close to the
limit
of Eq.$\,$\emminb\ and, given the uncertainties inherent in a
perturbative
expansion in $g$, it is fair to conclude that the search for
anomalies
in $Z \rightarrow 3\, \gamma$ transitions should be pursued with
vigour.
Consequently, I give more details on these decays.

Let $x_i\equiv 2\,E_i/M_{_Z}$ be the thrust of the individual
photons, with
$0\leq x_i \leq 1$ and $\sum_{i} x_i=2$. The Dalitz plot density is:
$$
{d\,\Gamma \over \Gamma \; d\,x_1\;d\,x_2\;d\,x_3}=
{5\over 34}\,\left[139\,\sum_{i} x_i^2(1-x_i)^2-18\,\prod_{i}
(1-x_i)\right]\,
\delta\left(\sum_{i} x_i-2\right)\; .
\eqn\eDalitz
$$
The single-photon energy distribution is:
$$
{1\over \Gamma}{d\,\Gamma \over d\,x}=
{x^3\,(3430-6210 x+2919 x^2)\over 102}\; .
\eqn\esingle
$$
The photons' angular distribution relative to the beam axis
(three entries per event) is:
$$
{d\,\sigma \over d\,\cos \theta}\propto
1-{77 \over 479}\,\cos^2(\theta)\; .
\eqn\esingle
$$
The angular distribution of the most energetic photon relative to the
beam direction is:
$$
{d\,\sigma \over  d\,\cos \Theta}\propto
1-0.1762\,\cos^2(\Theta)\; .
\eqn\ethrust
$$
The distribution in the angle between the normal to the three-photon
plane and the beam axis is:
$$
{d\,\sigma \over  d\,\cos \alpha}\propto
1+{163 \over 671}\,\cos^2(\alpha)\; .
\eqn\eplane
$$
All of these angular distributions are fairly uniform.

For a monopolar mass above the limits discussed in Section 3, the
ratio of five- to three-photon decay widths of the $Z$, as obtained
from Eqs.$\,$\ethreew,\efivew\ is a tiny number; the enhancement due
to the extra powers of $g$ is compensated by phase-space and
derivative-coupling factors. The three-photon decay of the $Z$
constitutes the best hope for a specific monopole-induced signature.

\vskip .2cm
\noindent{\caps 5. A digression on Monopolium}

Given the strength of the coupling of monopoles to $Z$'s and
$\gamma$'s
(and of dyons to $W$'s),
{\it monopolia}, tightly bound states of a monopole and its
antiparticle,
should exist. Are they the signatures to look for in $e^+e^-$
annihilation?
Do they affect our previous considerations? I argue that both
questions
ought to be answered negatively.

The monopolium bound state is not tractable by analogy with
positronium, or even charmonium. With a ``fine structure'' constant
$\alpha_g > 1$, one expects the constituent monopoles to be tightly
bound and highly non-relativistic.

Various authors [\rGKR] have computed monopolium masses by
``softening'' the potential with the introduction of a non-vanishing
extension of the monopoles, in ways  premonitory of the properties of
topological monopoles. They obtain mass estimates for the
ground-state monopolium some 200 times (!) lighter than the sum of
masses of the
constituents. While these results are to be taken {\it cum grano
salis,} they
convey the presumably correct feeling that monopolium states should
appear,
in $e^+e^-$ annihilations, well below the ``open monopole''
threshold.
Why are monopolia not the objects to hunt for?

Another calculation that cannot be confidently performed in analogy
with positronium is that of the lifetime of a monopolium state
against its decay into vector bosons. Recall that, for the
$3\,\gamma$ decay of the ($J^P=1^-$) orthopositronium ground state:

$$
\Gamma= (2 \,m_e) \times {(\pi^2- 9) \over 9\,\pi}\; \alpha_e^6\; .
\eqn\eortho
$$
Upon substitution of $\alpha_e$ for $\alpha_g$ one obtains a forcible
hint of a
foretold conclusion: monopolia ought to have a width much
larger than their mass. If so, they are not prominent peaks.

I expect the collective effect of wide monopolia to be describable at
the ``partonic'' level by the monopole--antimonopole loops we have
dealt with, in analogy with the QCD description of the energy-smeared
cross section for $e^+e^-$ annihilation into hadrons [\rHDR].
This completes the argument whereby monopolia ought to be irrelevant
in practice.

\vskip .2cm
\noindent{\caps 6. $W$-pair production in $e^+e^-$ collisions}

Dyons have a potential incidence on the $Z\,W^+\,W^-$ and
$\gamma\,W^+\,W^-$ Triple Gauge-Boson Vertices (TGVs) to be
thoroughly studied at LEP-II. To discuss these effects, it may be
useful to revamp one's souvenir of some known properties of the
Standard TGVs.

In the Standard Model, the tree-level TGVs are described by a
Lagrangian
$$
{\cal{L}}_{_{SM}}=-i\,{e\over s}
\left[\,\kappa_e\,
W_3^\nu\left(W^\dagger_{\mu\nu}W^\mu-
W_{\mu\nu}W^{\dagger\mu}\right)+
\kappa_m\;W_3^{\mu\nu}W^\dagger_\mu\,W_\nu\right]\; ,
\eqn\eLSM
$$
$$
W_3^\mu \equiv s\, A^\mu + c\, Z^\mu\; ,
\eqn\eWthree
$$

with $W_3^\mu$
the electrically-neutral isovector potential, and
$W_i^{\mu\nu}\equiv\partial^\mu W_i^\nu
-\partial^\nu W_i^\mu$.  In Eq.$\,$\eLSM,
$\kappa_e=\kappa_m=1$ are
 the non-anomalous ``electric'' and ``magnetic'' couplings of $W$'s
to $\gamma$ and $Z$. Radiative effects modify the coefficients
$\kappa_e,\,\kappa_m$ of the (dimension $d=4$) couplings of
${\cal{L}}_{_{SM}}$; the dispersive parts of these corrections are
akin to the charge form factor and magnetic anomaly ($g-2$) of an
elementary fermion.

Also in analogy with $g-2$, vertex corrections generate TGVs that are
describable by effective interactions not appearing in
${\cal{L}}_{_{SM}}$. The lowest dimension ($d=6$) of these TGVs is
$$
{\cal {O}}_{_W}\equiv W_3^{\mu\nu}
W^\dagger_{\nu\rho}\,W_\mu^\rho\; .
\eqn\eOW
$$
The corresponding effective Lagrangian is ${\cal L}_{_W}= a\, {\cal
{O}}_{_W}$,
with $a$ of order $e\alpha/(\pi\,M^2)$ and $M$ representative of the
mass of virtual particles. Effective operators such as ${\cal
{O}}_{_W}$ are particularly useful in discussing physics [\rLBW,
\rGavela] beyond the Standard Model.

As an example of non-standard effect, take a vector-like doublet of
conventional leptons $(E^0,E^-)_{_{L,R}}$ of degenerate mass
$M>M_{_Z}/2$.
This appendage affects the  couplings of Eq.$\,$\eLSM, entailing
modifications that (after renormalization) are of the form
$\Delta\kappa_e,\,\Delta\kappa_m\propto
e^3\,q^2/M^2$. The vertex correction illustrated in Fig.$\,$\fleptwo
a also induces effects that, to leading order in $1/M$, are described
by the addition to ${\cal{L}}_{_{SM}}$ of the effective interaction:
$$
{\cal{L}}_{_{E}}=-{i\over 240\,\pi^2\,M^2}\,
\left({e \over s}\right)^3\;
{\cal {O}}_{_W}\; .
\eqn\eLE
$$

Non-standard radiative effects may also induce CP-violating TGVs not
included in ${\cal{L}}_{_{SM}}$, two of which, of dimension $d=6$,
have been discussed in the literature\foot{The low-energy limits of
couplings of the form
$\widetilde{{\cal O}}_{\kappa}$ and
$\widetilde{\cal {O}}_{_W}$ describe electric dipole and magnetic
quadrupole moments of the $W$'s, and their electroweak
generalizations. Their standard values are of very high order of
perturbation theory, and negligible in practice.},
$\widetilde{{\cal O}}_{\kappa}\equiv
\widetilde{W}_3^{\mu\nu}W^\dagger_\mu\,W_\nu$,
the dual sibling of the Standard ``magnetic'' coupling of
Eq.$\,$\eLSM, and

$\widetilde{\cal {O}}_{_W}\equiv
\widetilde{W}_3^{\mu\nu}W^\dagger_{\nu\rho}\,W_\mu^\rho$,
the dual partner of ${\cal {O}}_{_W}$. A third possibility, a dual
counterpart of the ``electric'' ($\kappa_e$) coupling of
Eq.$\,$\eLSM, has not been discussed; and this for a good reason: its
coefficient at $q^2=0$ would be the magnetic-monopole charge of the
$W$.

Consider, at long last, the $O(ge^2)$ vertex corrections of
Figs.$\,$\fleptwo b,c, induced by our hypothetical dyon doublet
$(D^0,D^-)_{_{L,R}}$. We are interested in the lowest-dimension TGVs
characterizing this radiative effect or, more precisely,
in the interference between the standard $e^+e^-\rightarrow W^+W^-$
amplitude and
the amplitude induced by this correction.
For mass-degenerate dyons, the amplitudes of Figs.$\,$\fleptwo b,c
add to zero, since $D^-$ and $\bar D^0$ have opposite magnetic
charges (the ``Furry theorem'' stating the vanishing of this TGV is
in fact a consequence of the magnetic
analogue of charge-conjugation symmetry, and applies to all orders of
perturbation theory). To explore a possible non-vanishing effect, we
momentarily lift our restriction to mass-degenerate dyons.

 We saw in Section 2 that the dyon Feynman rules are not without
blemish and, indeed, their use to compute the amplitudes of
Figs.$\,$\fleptwo b,c  leads to troubles identical to the ones we
discussed there. To get out of this {\it cul de sac,} I simply
conjecture that a consistent set of computational rules for dyons
would result in an effective interaction analogous to the
unequal-mass generalization of Eq.$\,$\eLE\ but for the trading of
${\cal {O}}_{_W}$ for its dual,
$\widetilde{\cal {O}}_{_W}$, and the  substitutions of electric for
magnetic coupling constants. In support of this conjecture, recall
that there is no difficulty in ``taking the dual'' of an interaction
involving
$\vec E$ and $\vec B$
 fields rather than the vector potentials, essentially
all we suggest is  to make the replacement
$W_{\mu\nu}\rightarrow
\widetilde{W}_{\mu\nu}$ in Eqs.$\,$\eOW,\eLE\ while substituting
leptons for dyons\foot{Dyons also induce effects proportional to
$\widetilde{{\cal O}}_{\kappa}$, and a non-vanishing
``magnetic-charge radius'' of the $W$. We do not dwell on these
effects, which are of the same order of magnitude as the ones we do
discuss, and do not modify the conclusions.}.

Let the split dyon masses be $m_{-}^2=m^2+\mu^2$,
$m_0^2=m^2-\mu^2$. Following our conjecture and
recalling that $e\,g=2\,\pi\,n$, we may  compute the dyon-induced
TGV, for small $\mu^2/m^2$, as:
$$
{\cal{L}}_{_{D}}\simeq {i\,e\,n
\over 720\, \pi\,c\,s^2}\;
{\mu^2\over m^4}\;
\widetilde{B}^{\mu\nu}W^\dagger_{\nu\rho}\,W_\mu^\rho \; ,
\eqn\eLDyon
$$
with ${B}^{\mu\nu}$ the derivative fields of the $U(1)_{_Y}$
potential of Eq.$\,$\eGinv.
Behold! the operator in Eq.$\,$\eLDyon\ is not of the $\epsilon^{ijk}
W_i W_j W_k$ form of Eqs.$\,$\eLSM,\eOW; sacrosanct on grounds of the
gauge symmetry. The reason is that this operator is of dimension
$d=8$, in spite of its $d=6$ disguise.
The mass $m$, common to both (vector-like) dyons can be directly
ascribed to an invariant term in the Lagrangian. The mass $\mu$,
which splits them, must be proportional to a term such as
$\bar{D}M^0\langle\Phi\rangle$, and the Higgs field $\Phi$ accounts
for the extra field-dimensions.

The observability at LEP-II of  ``$T$-odd'' effects\foot{An
observable is $T$-odd if it  changes sign as the spin and momenta of
all particles are reversed.} induced by a CP-violating interaction
such as Eq.$\,$\eLDyon\
has been carefully investigated [\rBMDR].
Even for the most optimistic case ($\mu$ comparable to $m$, $m\sim 1$
TeV) the coefficient in Eq.$\,$\eLDyon\ is three orders of magnitude
below the level required to observe a $1\sigma$ effect in the
combined results of four LEP-II experiments, each stockpiling some
$10^4$ $W$ pairs [\rBMDR]. And, to make matters worse, the lower
limits on the mass of non-degenerate dyons are much stronger
than the limits we discussed in Section 3, for the same reason (a
breaking of the ``custodial'' $SU(2)$ symmetry) that a split $(t,b)$
quark pair contributes so significantly to radiative corrections.

The moral of this Section is that there appears to be no hope for
LEP-II, concerning monopoles and/or dyons, to improve on the
exclusion or discovery capabilities of LEP-I. This is an example of a
general result regarding the measurement of triple gauge-boson
vertices [\rGavela].
I have not explored the potential of higher-energy
$e^+e^-$ colliders.

\vskip .2cm
\noindent{\caps 7. Monopoles that are not point-like}

The very heavy composite monopoles that arise in Grand Unified
Theories are
certainly impossible to produce in current terrestrial laboratories.
It is conceivable,
however, that much lighter yet non-point-like monopoles exist, either
for
reasons that we have not at all fathomed, or because something like a
Grand
Unification actually takes place at an unsuspectedly low energy.
It is therefore interesting to ponder what the low-energy signatures
of
composite monopoles might be.

The grand-unified monopoles are classical or semi-classical solutions
to a gauge
theory.  They do not
correspond to a field appearing in the Lagrangian and, strictly
speaking,
they do not exist as virtual particles to be included in a quantum
loop
expansion, wherein only gauge bosons and other point-like particles
should play a role. Analogously, only quarks and gluons,
and perhaps their condensates, but not hadrons, ought to be used in
pure-QCD
perturbation theory (though useful alternative schemes do exist, such
as the chiral Lagrangians).

 Recall the foregone times when the quark model and QCD were not
established.
Imagine that the anomalous magnetic moment of muons and electrons
had been measured, by then, to a sufficient precision for the
hadronic
contribution to the photon's vacuum polarization, $\Pi (q^2)$, to
play a significant role.
To have theory agree with experiment one would need to estimate this
contribution. Use of a dispersion relation for $\Pi (q^2)$, and of
$\pi^+ \pi^-$ intermediate states with a form factor, $F_\pi (q^2)$,
describing the $\rho$ resonance, would result in an answer of the
correct magnitude, even though one would have violated the dictum of
the last paragraph: thou shalt not use but fundamental particles in a
quantum correction\foot{I am indebted to Andy Cohen for insisting on
this commandment, and on its exemptions.}.

For the rest of this Section, I imitate the procedure of the previous
paragraph, trading pions for extended monopoles. I also trade a
pion-like
form factor for a proton-like one, since
the monopole threshold should be above the monopolium states, like
the $p\bar p$ threshold is above the $\rho$ mass.
For the conventionally
defined electric and magnetic form factors at
threshold ($q^2=4 m^2$) I set a normalization $G_E=G_M\sim 0.5$,
the measured value for protons. Above threshold I let the
form factors decrease as in a dipole fit, with a slope that
corresponds to a mean square radius $\langle r^2\rangle=({\bar
\alpha}\,m)^{-2}$, with $\bar{\alpha}=1/40$.

With the use of conventional dispersion-relation techniques, one can
redo our previous calculations of monopole-mass limits and the
$Z\rightarrow 3\gamma$ width. The mass limits of Eqs.$\,$\emmina
--\emminc\ are weakened by a factor $\sim 3$, and
the $3\gamma$ branching ratio of Fig.$\,$\fwidth\ now reaches the
interesting level of 1 ppm
also for monopoles that are $\sim 3$ times lighter than the
point-like ones. These masses and mass limits are very low, some 200
GeV. Consistency of the extended-monopole weak-coupling picture
demands that the monopole-constituent gauge fields be much lighter
than the monopole.
For monopoles close to their lower mass limit, the novel gauge fields
should already have been found!

The moral of this Section's bantam exercise is not unexpected. Not
only are composite monopoles particularly difficult to produce, but
it should be much easier to find their ``constituents'' than any of
their indirect signatures.

\vskip .2cm
\noindent{\caps 8. Cosmological considerations}

Fossil remnants of any stable particle species may survive from the
time when our Universe was hot enough to sustain a thermal population
of its specimens, and monopoles are no exception. In a
monopole--antimonopole symmetric Universe, conventional Big-Bang
theory fixes the present relic monopole (and antimonopole) abundance
in terms of the monopole mass and $\sigma\cdot v$, the
velocity-weighed $M\bar M$ annihilation cross section\foot{An
excessive abundance of topological monopoles was one of the original
motivations for inflationary models [\rGuth]. We are exclusively
dealing in this section with point-like monopoles and assuming
inflation to have taken place at a temperature above the monopole
mass.}. To an admittedly suspicious lowest order in $\alpha_g$, the
total $\sigma\cdot v$ for annihilation into $ZZ$, $Z\gamma$ and
$\gamma\gamma$ is:
$$
\sigma\cdot v \simeq {4\,\pi \, \alpha_g^2 \over m^2\;c^4}\; ,
\eqn\esigma
$$
where we have used $m \gg M_{_Z}$, to be justified a posteriori. This
result also
applies to $D^+\,D^-$ or $D^0\bar{D}^0$ (dyon) annihilation, since
the additional
annihilation of dyons into $W$-pairs proceeds with a conventional
electroweak strength, and may be neglected.

Let $g$ and $g_{_{SM}}$ be the number of degrees of freedom of a
spin-$1/2$ monopole and of the complete Standard-Model zoo above the
weak scale ($g=4$, $g_{_{SM}}=427/4$). Let $\Omega=\rho/\rho_c$ be
the contribution of monopoles to the current universal energy
density, in current critical units $\rho_c\sim 2\,h^2\,10^{-29}$
g/cm$^3$, with the Hubble-constant  controversy  still raging between
$h\sim{1\over 2}$ and $h\sim 1$.
For ``cold relics'' such as monopoles, a conventional calculation
[\rKT] results in the following estimates of the decoupling
temperature $T_d$, and of $\Omega$:
$$
{T_d\over m}\simeq \left[A-{1\over 2} \ln (A)\right]^{-1}\;\;\;\;\;
(A\equiv 0.038\; g \;g^{-1/2}_{_{SM}}\;M_{_P}\,m\,\sigma\cdot v)
\eqn\eT
$$
$$
\Omega\,h^2 \simeq \left({1.1 \; 10^{12} \over 1\;\rm{TeV}}\right)\;
{m\over T_d} \;
                   {1 \over \sqrt{g_{_{SM}}} \, M_{_P} \, \sigma\cdot
v}\;\; ,
\eqn\eOm
$$
with $M_{_P}$ the Planck mass.

Equations \esigma-\eOm\ imply that, for monopoles or dyons to avert
``overclosing'' the Universe (contributing an $\Omega > 1$), their
mass must obey the restrictions:
$$
m < 8.7 \; n \; h \; \rm{PeV}\;\; \rm{[monopoles]\, ,}
\eqn\ecosmo
$$
$$
m < 6.1  \; n \; h \; \rm{PeV}\;\; \rm{[dyons]\, .}
\eqn\ecosdy
$$
A primordial monopole--antimonopole asymmetry would make the mass
limits more stringent.
Recall Eqs.$\,$\emmina,\emminb\ (and that 1 PeV $=10^3$ TeV) to
conclude that there is room for point-like monopoles only in a mass
interval of some four to five octaves.

\vskip .2cm
\noindent{\caps 9. Conclusions}

Magnetic monopoles are some of the most interesting particles that we
have not found.

The ``official'' non-elementary   topological monopoles are too heavy
to be made in the laboratory. Unsuspectedly light extensive monopoles
are also difficult to produce, much as a nucleus of anti-helium, due
to a severe ``form-factor'' suppression. Neither are their indirect
signatures, as we have discussed, a good way to look for them.

The virtual effects of point-like monopoles and dyons are observable
in experiments conducted at energies below their production
threshold. We have exploited the current data to set lower limits of
the order of 1/2 of a TeV on the masses of monopoles and degenerate
dyons, and the standard Big-Bang cosmology to set upper limits of the
order of 5000 TeV.

In the allowed mass interval, an unsuspectedly large $Z\rightarrow
\gamma\gamma\gamma$ branching ratio would be the most specific
currently accessible signature for the existence of magnetic poles.

\vskip .2cm
\noindent{\caps Acknowledgements.} I am indebted to Luis
Alvarez-Gaum\'e, Savas Dimopoulos,
Bel\'en Gavela, Howard Georgi, Gian Giudice, Shelly Glashow, Pilar
Hernandez and Ryan Rohm
for illuminating discussions, and I thank Andy Cohen for very many
useful,
constructive and instructive comments. I thank Elias Kiritsis, Tim
Morris and  Wolfgang Lerche for their exquisite patience with my
computer illiteracy. I am indebted to Bel\'en Gavela for a critical
reading of the typescript and I thank Samuel Ting, and
many other members of the L3 experiment, for discussions on the
measurement of photons at LEP.

\refout
\figout
\vfil\eject

\input epsf.tex
\nopagenumbers
\setbox2=\vbox to 2 truein{\epsfxsize= 7.5 truein\epsfbox
[0 0 612 792]{7273fg1.ps}}
\centerline{\box2}
\vskip -50pt
\vfil\eject

\setbox2=\vbox to 2 truein{\epsfxsize= 7.5 truein\epsfbox
[0 0 612 792]{7273fg2.ps}}
\centerline{\box2}
\vskip -50pt
\vfil\eject

\setbox2=\vbox to 2 truein{\epsfxsize= 7.5 truein\epsfbox
[0 0 612 792]{7273fg3.ps}}
\centerline{\box2}
\vskip -50pt
\vfil\eject

\setbox2=\vbox to 2 truein{\epsfxsize= 7.5 truein\epsfbox
[0 0 612 792]{7273fg4.ps}}
\centerline{\box2}
\vskip -50pt
\vfil\eject

\setbox2=\vbox to 2 truein{\epsfxsize= 7.5 truein\epsfboxssave
[0 0 612 792]{7273fg5.ps}}
\centerline{\box2}
\vskip -50pt
\end